\documentclass[prb,twocolumn,showpacs,floatfix,amsmath,amssymb,superscriptaddress]{revtex4-1}
\usepackage{amsfonts}
\usepackage{stmaryrd}
\usepackage{bbm}
\usepackage{mathrsfs}
\usepackage{tipa}
\usepackage{amssymb}
\usepackage{txfonts}
\usepackage{graphicx}
\usepackage{dcolumn}
\usepackage{epstopdf}
\usepackage[colorlinks,linkcolor=blue,urlcolor=blue,citecolor=blue]{hyperref}
\usepackage{multirow}
\usepackage{subfigure}
\usepackage{url}

\begin{document}
\newcommand*{\cm}{cm$^{-1}$\,}
\newcommand*{\Tc}{T$_c$\,}

\title{Optical spectroscopy and ultrafast pump-probe study on Bi$_2$Rh$_3$Se$_2$: evidence for charge-density-wave order formation}

\author{T. Lin}
\affiliation{International Center for Quantum Materials, School of
Physics, Peking University, Beijing 100871, China}

\author{L. Y. Shi}
\affiliation{International Center for Quantum Materials, School of
Physics, Peking University, Beijing 100871, China}

\author{Z. X. Wang}
\affiliation{International Center for Quantum Materials, School of
Physics, Peking University, Beijing 100871, China}

\author{S. J. Zhang}
\affiliation{International Center for Quantum Materials, School of
Physics, Peking University, Beijing 100871, China}

\author{Q. M. Liu}
\affiliation{International Center for Quantum Materials, School of
Physics, Peking University, Beijing 100871, China}

\author{T. C. Hu}
\affiliation{International Center for Quantum Materials, School of
Physics, Peking University, Beijing 100871, China}

\author{T. Dong}
\affiliation{International Center for Quantum Materials, School of Physics, Peking University, Beijing 100871, China}

\author{D. Wu}
\affiliation{International Center for Quantum Materials, School of
Physics, Peking University, Beijing 100871, China}

\author{N. L. Wang}
\email{nlwang@pku.edu.cn}
\affiliation{International Center for Quantum Materials, School of Physics, Peking University, Beijing 100871, China}
\affiliation{Collaborative Innovation Center of Quantum Matter, Beijing 100871, China}

\begin{abstract}

The parkerite-type ternary chalcogenide Bi$_2$Rh$_3$Se$_2$ was discovered to be a charge density wave (CDW) superconductor. However, there was a debate on whether the observed phase transition at 240 K could be attributed to the formation of CDW order. To address the issue, we performed optical spectroscopy and ultrafast pump-probe measurements on single crystal samples of Bi$_2$Rh$_3$Se$_2$. Our optical conductivity measurement reveals clearly the formation of an energy gap with associated spectral change only at low energies, yielding strong evidence for a CDW phase transition at 240 K. Time resolved pump-probe measurement provides further support for the CDW phase transition. The amplitude and relaxation time of quasiparticles extracted from the photoinduced reflectivity show strong enhancement near transition temperature, yielding further evidence for the CDW energy gap formation. Additionally, a collective mode is identified from the oscillations in the pump-probe time delay at low temperature. This mode, whose frequency decreases gradually at elevated temperature, could be naturally attributed to the amplitude mode of CDW state.
\end{abstract}

\maketitle
\section{introduction}

Charge-density wave (CDW) and superconductivity are two important and closely linked broken symmetry states in solids. There has been tremendous interest in the interplay between these two states in condensed matter physics. The topic has motivated extensive exploration of new materials showing coexistence or competition between the two different instabilities. The discovery of a coexistence of superconductivity and a structural phase transition in Bi$_2$Rh$_3$Se$_2$ represents a new progress \cite{Sakamoto2007} on this subject and offers new opportunity to study the CDW instability and its interplay with superconductivity.

Bi$_2$Rh$_3$Se$_2$ belongs to a quasi-two dimensional parkerite-type ternary chalcogenides A$_2$M$_3$X$_2$ (A = Sn, Pb, In, Tl, and Bi; M = Co, Ni, Rh, and Pd; X = S and Se) composed of sheets containing one-dimensional M-M chains \cite{Sakamoto2007,NATARAJAN1988215,Weihrich2007}. Sakamoto et al. reported that Bi$_2$Rh$_3$Se$_2$ was a new superconducting compound with a transition temperature T$_c \sim$ 0.7 K. Intriguingly, the compound exhibits a phase transition at about 240 K. Based on resistivity, magnetic susceptibility, specific heat measurement, thermoelectric power, thermal expansion, and low-temperature x-ray measurements, they identified the phase transition at 240 K as a CDW order \cite{Sakamoto2007}. Following this work, Kaluarachchi et al. \cite{Kaluarachchi2015} found that the isostructural compound Bi$_2$Rh$_{3.5}$S$_2$ has a higher superconducting transition temperature  T$_c \sim$ 1.7 K, though the stoichiometric  Bi$_2$Rh$_3$S$_2$ is not superconducting. A first order phase transition at 165 K was also found for Bi$_2$Rh$_3$S$_2$. The simultaneous observation of superconductivity and CDW order has brought a new perspective to the research in this field. However, the subsequent pressure and selected-area electron diffraction studies on Bi$_2$Rh$_3$Se$_2$ by Chen et al.\cite{Chen2014} indicated that the resistivity anomaly at 240 K shifted to higher temperature with increasing pressure, which is unusual for a conventional CDW transition. They argued that the phase transition at 240 K is not a CDW transition, but a purely structural phase transition with a symmetry reduction from a high-symmetry C-centered monoclinic lattice to a low-symmetry primitive one below the transition temperature \cite{Chen2014}.

It is crucial to understand the nature of the phase transition in those compounds because it is an essential step towards understanding the properties of compounds and the possible connection to the superconductivity. Up to now,
there is no spectroscopic experiment on those compounds. It is well known that most of the CDW states are driven by the nesting topology of Fermi surfaces, i.e. the matching of sections of FS to others by a wave vector \textbf{q} = 2\textbf{k}$_F$, where the electronic susceptibility has a divergence.
A single-particle energy gap opens in the nested regions of the Fermi surfaces at the transition, which leads to the lowering of the electronic energies of the system. Simultaneously, the phonon mode of acoustic branch becomes softened to zero frequency at \textbf{q} = 2\textbf{k}$_F$ as a result of electron-phonon interaction, leading to structural distortion \cite{densitywave}. The formation of an energy gap below the transition has been generally taken as a characteristic feature of CDW order. On the contrary, a purely structural phase transition, if irrelevant to a CDW order, would lead to an entirely different band structure, resulting in a spectral change over broad energy scale rather than only in the low energies. Such broad-energy spectral change across the phase transition was demonstrated in a number of materials such as BiNi$_2$As$_2$ \cite{PhysRevB.80.094506}, IrTe$_2$ \cite{Fang2013}, RuP \cite{PhysRevB.91.125101}. Additionally, CDW also has collective excitations, being referred to as an amplitude mode (AM) and a phase mode. The amplitude mode involves the ionic displacement and has a finite energy even at \textbf{q}=0 limit, which could be identified by the ultrafast pump-probe experiment \cite{PhysRevLett.83.800,PhysRevB.66.041101,PhysRevLett.101.246402,PhysRevB.78.201101,1347-4065-46-2R-870}. Furthermore, the formation of a CDW energy gap would also impede the relaxation time of photoexcited quasiparticles which could be also probed from the pump-probe measurement.

In this work, we performed optical spectroscopy and ultrafast pump-probe measurements on single crystal samples of Bi$_2$Rh$_3$Se$_2$. Our measurement reveals clearly the formation of an energy gap with associated spectral change only at low energy, yielding strong evidence for a CDW phase transition at 240 K. The opening of the energy gap removes most of the free carrier spectral weight and causes a dramatic reduction in the carrier scattering rate. Ultrafast pump-probe measurement reveals a significant change of the photoinduced reflectivity near the phase transition temperature. A strong enhancement of the amplitude and relaxation time of photoinduced carriers is extracted at the phase transition temperature, also yielding evidence for the CDW energy gap opening. Moreover, the time resolved measurement demonstrates presence of a strong oscillation at low temperature, which becomes damped gradually at elevated temperature. The temperature dependence of the oscillation suggests that it comes from the amplitude mode of CDW collective excitations.

\section{Results and discussion}

The single crystal samples of Bi$_2$Rh$_3$Se$_2$ were synthesized by self flux method. High-purity Bi, Rh, and Se elements with a molar ratio of Bi:Rh:Se=2:3:2 were mixed,  placed in an alumina crucible, and sealed in a silica ampule filled with argon gas. The sealed ampule was heated to 1050 $^0$C and held for 5 h, then slowly cooled with the rate of 2 $^0$C/h to 750 $^0$C. At the final temperature, the mixture was decanted using a centrifuge.

\begin{figure}[htbp]
\centering
\includegraphics[width=0.35\textwidth]{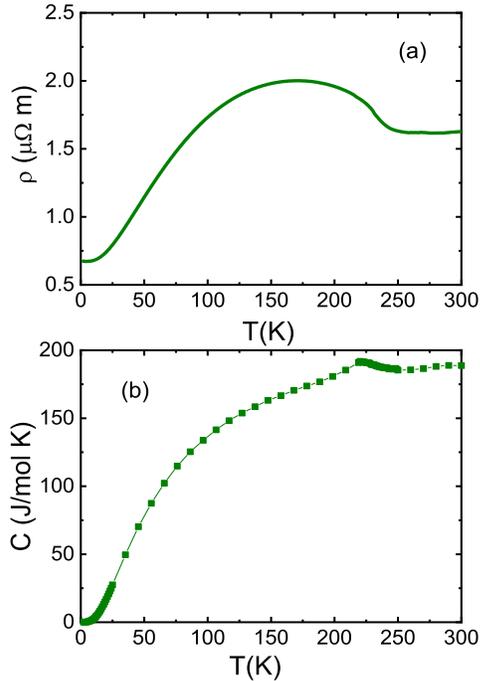}
\caption{(a)Temperature dependent resistivity, (b) temperature dependent of specific heat of Bi$_2$Rh$_3$Se$_2$. A phase transition is evident near 240 K. }\label{Fig:Res-SH}
\end{figure}

The temperature dependent resistivity was measured by a standard four-probe method. The specific heat was measured by using the relaxation method. Both were performed in a Quantum Design physical property measurement system (PPMS). Figure \ref{Fig:Res-SH} (a) and (b) show the temperature dependence of resistivity and specific heat between 1.8-300 K, respectively. Similar to previous reports \cite{Sakamoto2007,Chen2014}, a phase transition is clearly observed starting from 240 K.

\begin{figure}[htbp]
  \centering
\includegraphics[width=0.35\textwidth]{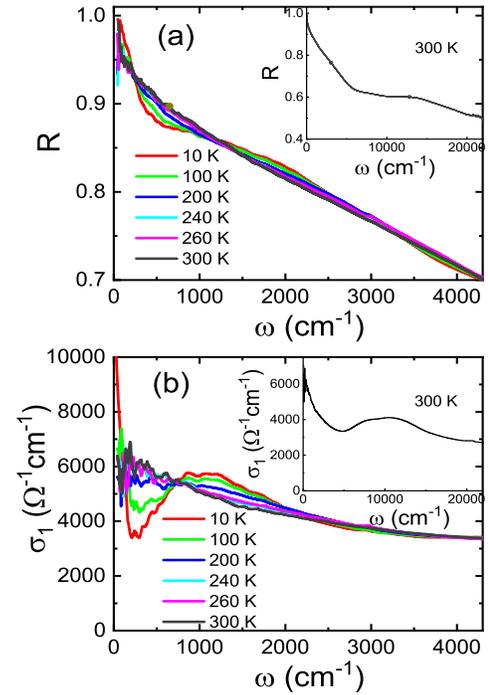} 
  \caption{The temperature dependent (a) reflectivity $R(\omega)$ and (b) optical conductivity $\sigma_1(\omega)$ for Bi$_2$Rh$_3$Se$_2$ single crystal below 4300 \cm, respectively. The insets of (a) and (b) present $R(\omega)$ and $\sigma_1(\omega)$ at 300 K over a broad frequency range, respectively.}\label{Fig:ref-cond}
\end{figure}

The in-plane reflectivity $R(\omega)$ was measured by the Fourier transform infrared spectrometer Bruker 80V in the frequency range from 40 to 25 000 \cm.
The room temperature $R(\omega)$  over a broad frequency range is displayed in the inset of Fig.\ref{Fig:ref-cond} (a). The spectrum shows typically metallic frequency response: $R(\omega)$ has high values at low frequencies and approaches unit at zero frequency limit. A roughly linear-frequency dependent reflectivity is seen below 6000 \cm. The behavior is similar to high-T$_c$ cuprate superconductors, reflecting an overdamped behavior of carrier scattering.
The main panel of Fig.\ref{Fig:ref-cond} (a) shows the reflectivity below 4300 \cm at several selected temperatures. Above
240 K, $R(\omega)$ increases monotonically as frequency deceases and the low energy part increases slightly upon cooling, both of which belong to simple metallic behaviors. With temperature further decreasing, a pronounced dip structure appears roughly near 600 \cm, which yields strong evidence for the formation of a charge gap in the vicinity of Fermi level due to the development of the CDW order. This dip feature grows more dramatic as temperature decreases, indicating the continuous enhancement of the CDW gap. In the meantime, the low energy reflectivity gets even higher than the values in the high temperature phase.

The real part of optical conductivity $\sigma_1(\omega)$ was derived from $R(\omega)$ through Kramers-Kronig transformation, as shown in Fig.\ref{Fig:ref-cond} (b).
The Hagen-Rubens relation was used for the low energy extrapolation of $R(\omega)$. For the high frequency extrapolation we have employed the x-ray atomic scattering functions \cite{PhysRevB.91.035123}. The main panel of Fig.\ref{Fig:ref-cond} (b) shows $\sigma_1(\omega)$  below 4300 \cm at different temperatures, the inset shows $\sigma_1(\omega)$ over broad energy scale at room temperature. Above the transition temperature, the optical conductivity exhibits clearly a Drude peak at low frequency. Its broad width indicates a large scattering rate $\gamma$ of the free carriers. Upon entering the low temperature phase, the spectral weight of the Drude peak was substantially removed and transferred to higher energies to form a broad peak centered around 1000 \cm. The feature becomes more prominent as temperature decreases. In fact, those are expected spectroscopic features for the density wave condensate, because either charge or spin density wave condensate has a so-called "case-I" coherence factor which would cause a sharp rise in the optical conductivity spectrum just above the energy gap\cite{densitywave}. Here, we can take the central position of the peak as the upper limit energy scale of the CDW gap. Then, we get the value of $2\Delta/k_B T_{CDW}\sim$6, a value being larger than the weak-coupling BCS theory but not uncommon for a density wave phase transition. We emphasize that those features are dramatically different from a purely structural phase transition. For a structural phase transition irrelevant to CDW order, the band structure at low temperature phase could be entirely different from the high temperature phase. That would result in a sudden spectral change over broad energy scale, which has been observed in a number of materials such as BiNi$_2$As$_2$ \cite{PhysRevB.80.094506}, IrTe$_2$ \cite{Fang2013}, RuP \cite{PhysRevB.91.125101} across the structural phase transitions.

\begin{figure}[htbp]
  \centering
\includegraphics[width=0.35\textwidth]{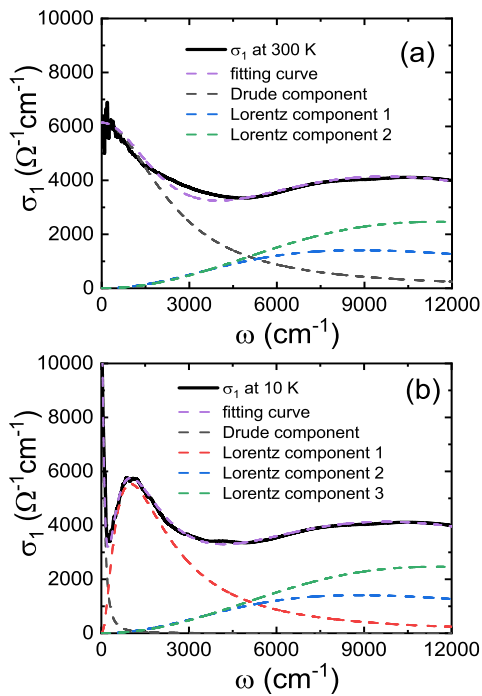} 
  \caption{The frequency dependent optical conductivity $\sigma_1(\omega)$ at 300 K (a) and 10 K (b), together with a Drude-Lorentz fit. The fitting parameters of high frequency Lorentz components are kept unchanged, therefore, the removed spectral weight from the Drude component at high temperature is transferred to the peak centered near 1000 \cm.}
  \label{Fig:cond-fit}
\end{figure}

To estimate the gapped spectral weight, we employ a Drude-Lorentz model to fit the optical conductivity,
\begin{equation}
\sigma_1(\omega)= {\frac{\omega_p^2}{4\pi}}{\frac{\Gamma_D}{\omega^2+\Gamma_D^2}}+ \sum_{j}{\frac{S_j^2}{4\pi}}{\frac{\Gamma_j\omega^2}{(\omega_j^2-\omega^2)^2+\omega^2\Gamma_j^2}}.
\label{chik}
\end{equation}
The Drude term describes the response of itinerant carriers, while the Lorentz terms stand for interband transitions and excitations across energy gaps. Here, for simplicity, we use only one Drude component to estimate the spectral weight of itinerant carriers. Figure \ref{Fig:cond-fit} shows the fitting results at two representative temperatures, 300 K and 10 K. We find that the low frequency spectrum at 300 K can be approximately reproduced by one Drude component. At 10 K, the removed spectral weight forms a peak at about 1000 \cm ($\sim$ 120 meV), an indication of energy gap formation. Since the fitting parameters of high frequency Lorentz components are kept unchanged, the spectral weight transfer occurs only in the low energy scale. It is worthy noting that a sharp and narrow Drude peak remains in the low temperature state, which agrees well with the high reflectivity near zero frequency, indicating that the Fermi surfaces are partially gapped by the CDW phase transition. The number of lost free carriers could be estimated by the variation of plasma frequency $\omega_p\sim\sqrt{n/m^*}$, where $n$ and $m^*$ represent the number and effective mass of free carriers respectively. From the above decomposition, we find that the plasma frequency $\omega_p$ varies roughly from 2.8$\times10^4$ \cm at room temperature to 8$\times10^3$ \cm at 10 K, which indicates that over 90 \% of free carriers are removed by the opening of CDW energy gap. Meanwhile, the scattering rate $\Gamma_D$ decreases violently from 2540 \cm to 93 \cm, as been evidenced by the narrowing of Drude peak, which makes the compound to have even lower \emph{dc} resistivity at low temperature despite of the substantial carrier density loss. Actually, the plasma frequency associated with the Drude component spectral weight could also be estimated by $\omega_p^2=8\int_0^{\omega_c}\sigma_1(\omega)d\omega$, where $\omega_c$ is the cut-off frequency of the Drude component. Taking the location of the conductivity minimum as the cutoff frequency, being about 4800 \cm for 300 K and 290 \cm for 10 K, where the balance between the tails of Drude and Lorentz components are roughly taken into account, we obtained the same values of plasma frequencies as that from the above decomposition of the spectral weight.

Further support for the formation of CDW order can be obtained from our ultrafast pump probe measurement, which has been proven to be particularly useful in detecting both of the single particle excitations across small energy gaps\cite{PhysRevLett.82.4918,PhysRevLett.104.027003,PhysRevB.84.174412,Chen2014a,PhysRevB.75.115120} and collective modes relevant to lang range ordering \cite{PhysRevLett.101.246402,Albrecht1992,PhysRevLett.111.057402}. We used a Ti:sapphire oscillator as the light source for both pump and probe beams, which can produce 800 nm pulsed laser at 80 MHz repetition. The 100 femtosecond time duration of the laser pulses enables ultrashort time resolved measurement. The fluence of the pump beam is about 6.4 $\mu J/cm^2$, and the fluence of the probe beam is ten times lower. In order to reduce the noise caused by stray light, the pump and probe pulses were set to be cross polarized and an extra polarizer was mounted just before the detector.

\begin{figure*}[hbtp]
  \centering
\includegraphics[width=18cm]{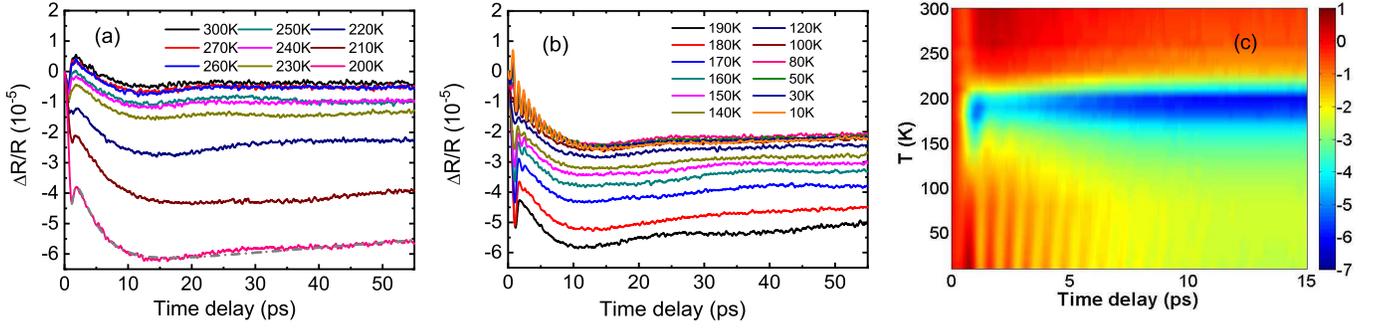}\\
  \caption{(a) $\Delta R/R$  in the temperature range of 200 - 300 K. The absolute value of $\Delta R/R$ increases with decreasing temperature, reaching the maximum at 200 K. (b) $\Delta R/R$  in the temperature range of 10 - 200 K. The absolute value of $\Delta R/R$ decreases with further decreasing temperature. The grey dash-dot curve on top of the data at 200 K is the fitting curve from equation (2). (c) The intensity plot of $\Delta R/R$ below 15 ps in different temperatures. An intensity oscillation is clearly seen at low temperature, suggesting presence of a collective mode.  The periodic time of the oscillation increases upon temperature increasing, suggesting a decrease of mode frequency at elevated temperature. }
  \label{Fig:DeltaR}
\end{figure*}

The photoinduced reflectivity change $\Delta R/R$ as a function of time delay at different temperatures are displayed in Fig.\ref{Fig:DeltaR} (a) and (b), respectively. Overall, the signal levels change quickly below 240 K. With temperature decreasing, the absolute value of $\Delta R/R$ increases and reaches the maximum near 200 K (Fig.\ref{Fig:DeltaR} (a)), then the amplitude drops with further decreasing temperature (Fig.\ref{Fig:DeltaR} (b)). The intensity plot for the time delay below 15 ps for different temperatures is shown in Fig.\ref{Fig:DeltaR} (c). We  observe three distinct decay processes: a fast relaxation time with negative amplitude value, a longer relaxation time with positive amplitude and another very long relaxation time with negative amplitude. Indeed, the reflectivity change could be well reproduced by three exponential decays,
\begin{equation}
\Delta R/R= A_1 exp(-t/\tau_1) +A_2 exp(-t/\tau_2)+A_3 exp(-t/\tau_3),
\label{chik}
\end{equation}
where $A_i$ (i=1,2,3) represents the amplitude of the photoinduced reflectivity change and $\tau_i$ stands for the relaxation time of the decay channel. As an example, we show the fitting curve for the data at 200 K (the bottom curve in Fig.\ref{Fig:DeltaR} (a)) by this formula. The amplitudes $A_1$, $A_2$ and $A_3$ have the same order of magnitude, whereas the relaxation times of the three decay processes are dramatically different: $\tau_1$ is sub-picosecond, $\tau_2$ is a few picoseconds, and $\tau_3$ are several hundred picoseconds. At 200 K, for example, $A_1$=-1.70 and $\tau_1$=0.6 ps, $A_2$=3.82 and $\tau_2$=3.8 ps, and $A_3$=-6.47 and $\tau_3$=350 ps, respectively.

Presence of rapid and slow decay dynamics after excitation has been observed in many systems \cite{Kumar2013,Luo2012,Tomeljak2009}. In general, the number of photoexcited hot electrons (or quasiparticles) at zero time delay is related to the amplitude of $\Delta R/R$. Those excited high-energy hot electrons release and transfer their energy to lattice through the emission of longitudinal optical phonons, and those optical phonons further decay into longitudinal acoustic phonons via anharmonic interactions. The sub-picosecond decay would be mainly attributed to quasiparticle relaxation via the electron-phonon thermalization, the several to several hundred picosecond decay processes would be related to the lattice energy loss via phonon population decay (inelastic scattering) or dephasing (elastic scattering) \cite{Luo2012,Hase2005}. The energy loss from the excited hot spot to the ambient environment would take even longer time.

\begin{figure}[htbp]
  \centering
\includegraphics[width=0.35\textwidth]{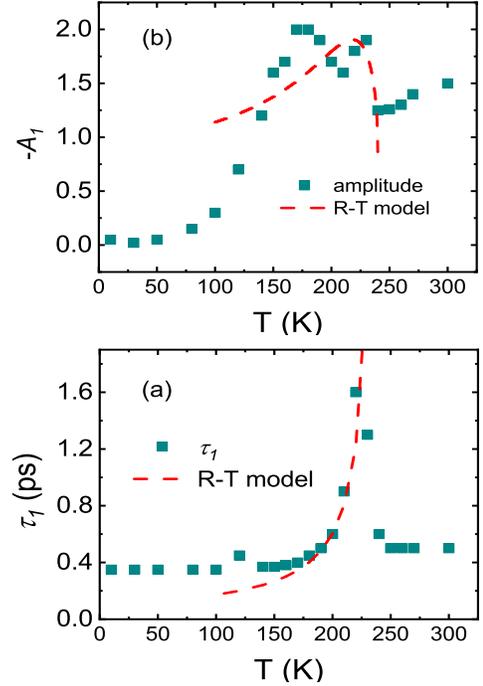} 
  \caption{The amplitude A$_1$ and relaxation time $\tau_1$ of the fast decay process extracted from the fitting of equation (2) to the experimental data of $\Delta R/R$ at different temperatures. The red dash curves are fitting curves from equation (3) and (4) below the phase transition temperature.}\label{Fig:pump-probe-fit}
\end{figure}

The substantial change of $\Delta R/R$ below 230$\sim$240 K reflects the phase transition. From the fitting of equation (2) to $\Delta R/R$, we find that the $A_i$ and $\tau_i$  (i=1,2,3) parameters for all three decay processes change with temperatures, particularly near the phase transition temperature. Such situation was observed in other materials before, for example, in spin density wave compound CaFe$_2$As$_2$ \cite{Kumar2013}. As our focus here is on the issue of whether the structural phase transition near 240 K is related to the CDW order, we shall limit our attention to the quasiparticle relaxation in sub-picosecond decay channel. Figure \ref{Fig:pump-probe-fit} shows A$_1$  and $\tau_1$ extracted from the fitting of equation (2) to $\Delta R/R$ as a function of temperature. An abrupt increase of A$_1$ is observed at the transition temperature with a broad peak-like structure at lower temperature, and simultaneously, a divergence in $\tau_1$ is seen near the transition temperature. Those features represent ultrafast spectroscopy evidence for an energy gap opening, as we shall explain below.

It is well known that the energy gap formation has a significant effect on the decay dynamics the photoexcited quasiparticles, which was described by the phenomenological Rothwarf-Taylor (R-T) model \cite{RT1967}. This model was initially established to explain the ultrafast relaxation dynamics of superconductors, but was proved to be applicable for a wide range of metallic systems with gap opening in the density of states. It proposes that the high energy phonons emitted by the recombination of photoinduced quasiparticles across an energy gap will introduce a bottleneck effect to the relaxation. That is, the depletion of states near the Fermi level would significantly impede the relaxation of the photoinduced quasiparticle. In the small photo excitation limit, the R-T model relates the density of thermally activated quasiparticles, $n_T$, to the measured transient reflectivity amplitudes $A(T)$ as, $n_T \propto \mathcal{A}^{-1}-1$, where $\mathcal{A}=A(T)/A(T\rightarrow 0)$. Assuming standard form of thermally activated quasiparticle density $n_T \propto \sqrt{T\Delta(T)}exp[-\Delta(T)/2T]$ and a BCS-like gap of the form $\Delta(T)=\Delta(0)\sqrt{1-T/T_c}$, the amplitude of photoinduced reflectivity signal is given by \cite{PhysRevB.59.1497}
\begin{equation}\label{Eq:1}
A(T) \propto {\frac{\Phi/(\Delta(T)+k_BT/2)}{1+\gamma\sqrt{k_BT/\Delta(T)}exp[-\Delta(T)/k_BT]}},
\end{equation}
where $\Phi$ is the pump fluence and $\gamma$ is a fitting parameter. The equation describes an increase in the photoexcited quasiparticle density due to the decreasing gap value and corresponding enhanced phonon emission during the relaxation. As the gap closes at transition temperature, more and more low-energy phonons become available for reabsorption, a quasi-divergence in the relaxation time is resulted\cite{PhysRevB.59.1497},
\begin{equation}\label{Eq:1}
\tau \propto {\frac{ln(g+exp(-\Delta(T)/k_BT))}{\Delta(T)^2}},
\end{equation}
where g is a fitting parameter. Indeed, the sudden change of $A_1$ and $\tau_1$ near the phase transition temperature could be captured by equations (3) and (4) of R-T model, as shown in Fig. \ref{Fig:pump-probe-fit}. There exists some deviation at lower temperatures. It might be linked to the simultaneous change in $A_2$, $\tau_2$ and $A_3$ and $\tau_3$, which was ignored in the above analysis. Additionally, we expect that the R-T model, which is appropriate near the phase transition, would become less applicable at temperature far below the phase transition. Nevertheless, the sudden increase in $A_1$ and quasi-divergence in $\tau_1$ illustrate unambiguously the appearance of an energy gap, yielding further evidence for the CDW phase transition.

\begin{figure}[htbp]
  \centering
\includegraphics[width=0.35\textwidth]{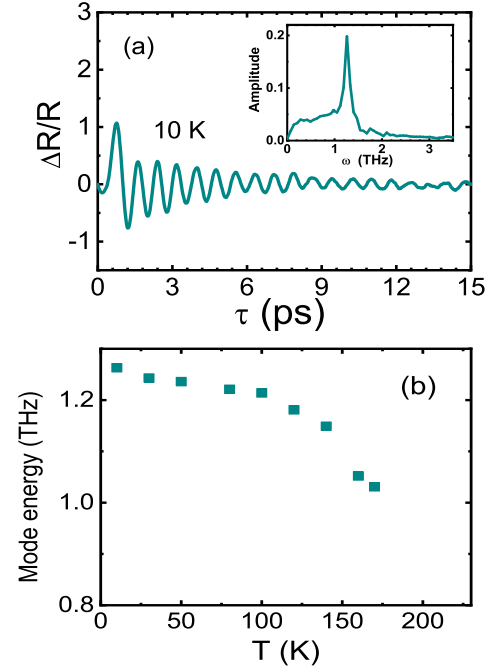} 
  \caption{(a) $\Delta R/R$ with pump-probe time delay within 15 ps at 10 K. The decay background is subtracted. The inset shows the mode in frequency domain after the Fourier transformation. (b) The extracted mode frequency from the oscillations in pump-probe measurement as a function of temperature.}\label{Fig:CDWmode}
\end{figure}

On the other hand, CDW also has collective excitations, being referred to as amplitude mode and phase mode. The amplitude mode involves the ionic displacement and has a finite energy even at \textbf{q}=0 limit, usually at the energy level of terahertz frequency, which could be identified by the ultrafast pump-probe experiment \cite{PhysRevLett.83.800,PhysRevB.66.041101,PhysRevLett.101.246402,PhysRevB.78.201101,1347-4065-46-2R-870}.
Indeed, the pump-probe signals exhibit pronounced oscillations at low temperature, as seen clearly in Fig. \ref{Fig:DeltaR} (b) and intensity plot of Fig. \ref{Fig:DeltaR} (c). It is easy to find that the periodic time of the oscillation increases upon temperature increasing, suggesting a decrease of mode frequency. In order to analyze the oscillatory component quantitatively, we subtract the exponential fitting part, then perform fast Fourier transformation of the residual part to get the mode frequency. As an example, Fig. \ref{Fig:CDWmode} (a) shows the $\Delta R/R$ data at 10 K as a function of time delay within 15 ps with the decay background subtracted. The inset shows the mode in frequency domain after the Fourier transformation. The result for different temperature is plotted in Fig. \ref{Fig:CDWmode} (b). The mode frequency of the oscillation drops as temperature increases. In CDW compound, the collective amplitude mode of CDW condensate usually has higher signal strength than phonon modes and behaves as an order parameter as a function of temperature. In some CDW materials, one indeed observes the disappearance of CDW amplitude mode precisely at the CDW transition temperature \cite{PhysRevLett.118.107402}. But quite often, the oscillations are heavily damped at elevated temperature and could not be resolved before reaching the transition temperature. This is also the case for the present compound. Thus, judging from the signal level and the temperature dependent trend, we can attribute the observed mode to the CDW amplitude mode. The mode frequency is about $\Omega_A$=1.25 THz at T=0 K, which is among the commonly observed energy scales for CDW order. In fact, the CDW amplitude mode frequency at the Brillouin zone center is known to be related to the phonon frequency of acoustic branch $\omega_{2k_F}$ at wave vector \textbf{q} = 2\textbf{k}$_F$ above $T_{CDW}$ and the electron-phonon coupling constant $\lambda$ by $\Omega_A\sim \lambda^{1/2}\omega_{2k_F}$ \cite{densitywave}. Unusual CDW mode frequency was observed only for certain compound with peculiar CDW wave vector \textbf{q} = 2\textbf{k}$_F$, for example, in LaAgSb$_2$ with extremely small CDW wave vector 2\textbf{k}$_F$ or unusual long lattice modulation period in real space \cite{PhysRevLett.118.107402}.

\section{Summary}
To summarize, we have utilized infrared spectroscopy and ultrafast pump-probe measurement to investigate the charge and coherent dynamics of the Bi$_2$Rh$_3$Se$_2$ single crystals, in an effort to address whether the phase transition at 240 K is a CDW order or a purely structural transition. Our optical spectroscopy measurement reveals clearly the formation of an energy gap with associated spectral change only at low energies, yielding strong evidence for a CDW phase transition at 240 K. The formation of the energy gap removes most part of the free carrier spectral weight. Time resolved pump-probe measurement provides further support for the CDW phase transition. The amplitude and relaxation time of quasiparticles extracted from the photoinduced reflectivity show strong enhancement near transition temperature, yielding further evidence for the CDW energy gap formation. Additionally, a collective mode is identified from the oscillations in the pump-probe time delay at low temperature. This mode, whose frequency decreases gradually at elevated temperature, is suggested to be the amplitude mode of CDW condensate state.

\begin{center}
\small{\textbf{ACKNOWLEDGMENTS}}
\end{center}

This work was supported by National Natural Science Foundation of China (No. 11888101), the National Key Research and Development Program of China (No. 2017YFA0302904, 2016YFA0300902).

\bibliographystyle{apsrev4-1}
  \bibliography{BiRhSe}

\end{document}